\documentclass{article}
\usepackage[utf8]{inputenc}
\usepackage[table,xcdraw]{xcolor}
\usepackage{epsfig}  \usepackage{float}   \usepackage{csquotes}
\usepackage{graphics} \usepackage{graphicx}        
\usepackage{subfigure}          %\usepackage[
 \usepackage[citestyle=authoryear-comp,natbib=true,backend=bibtex]{biblatex}
\usepackage{fancyheadings}          \usepackage[utf8]{inputenc}
\usepackage[english]{babel}
\usepackage{float}          
\usepackage{times}       
\usepackage{amsmath,amsfonts,amssymb,amscd,amsthm,xspace}         
\usepackage{algorithmic}         
\usepackage{algorithm2e}        
\usepackage{verbatim}        
\usepackage{appendix}         
\usepackage{hyperref}
\usepackage{enumitem}
\title{Data Markets to support AI for All: Pricing, Valuation and Governance}
\author{Ramesh Raskar, Praneeth Vepakomma, Tristan Swedish, Aalekh Sharan*\\
Massachusetts Institute of Technology\\
Cambridge, MA 02139, USA \\
\texttt{\{raskar,vepakom,tswedish\}@mit.edu}
}
\date{}

\begin{document}

\maketitle
\section{Introduction}
We discuss a data market technique based on intrinsic (relevance and uniqueness) as well as extrinsic value (influenced by supply and demand) of data. For intrinsic value, we explain how to perform valuation of data in absolute terms (i.e just by itself), or relatively (i.e in comparison to multiple datasets) or in conditional terms (i.e valuating new data given currently existing data).
\section{Motivation for creating Data Markets}

%There is no mastermind in directing the coordination of a market. The ‘invisible hand of the market’, with its only weapon being price signals is able to incentivize collaboration amongst the various entities involved in the market by thinking selfishly of maximizing their own gains. 

\textbf{AI will benefit from Liquid Markets:} %When these thoughts are extended to the AI economy, it is apparent that these liquid markets do not exist today.
Data is increasingly concentrated in large firms. For startups, and small organizations it is increasingly difficult to compete as the lack of availability of data can stymie any and all efforts to build better machine learning algorithms. Algorithmic capability indeed increases with the availability and quality of data. One way to tackle this is the marketplace approach. By creating conditions such that data, the raw material for AI, can be bought and sold with security, privacy and consent safeguarded, specialized niches will be created and firms will be able to tackle a smaller subset of the problem.  A precondition for this large scale collaboration to occur is the existence of liquid markets at various steps of the value chain.

\textbf{Example:} Consider a diagnostic healthcare company aiming to do an acquisition of labeled X-ray images from various hospitals for developing state of the art diagnostics. %The prospective data providers in this case would be patients and healthy subjects with varied signatures of health data and conditions. 
The key problem in such a setting is: ``How can value of datasets from each hospital be estimated to decide their price?" The data for some hospitals can belong to unique health traits and demographics and can be very valuable for the diagnostic use-case of the company while data from some other hospitals may be of a relatively much lower value.\\
A key problem therefore is that obtaining large amounts of diverse, yet useful data costs a lot of resources. There are also diminishing returns at some point when additional data does not improve the capabilities of the algorithm, if additional data is not acquired intelligently in a cost-effective manner.

\section{Data Valuation for AI}
\textbf{Absolute, relative or conditional data purchase:}
Another required facet to setting up a data market is to build  a capability to perform valuation of data in absolute terms (i.e just by itself), or relatively (i.e in comparison to multiple datasets) or in conditional  terms (i.e valuating new data given currently existing data). \par \textbf{Intrinsic or extrinsic data valuation:} Any of these data valuation use-cases can be performed via intrinsic factors of evaluation such as based on quality of information within the dataset or via extrinsic factors of evaluation such as based on demand-supply, market economics, game theoretic mechanisms and speculative market forces or via a combination of both as in \citep{koutris2015query,koutris2013toward,balazinska2013discussion,deep2017qirana,li2014theory,zheng2019arete,zheng2017trading,zheng2017online}. \par \textbf{Goal dependent or independent data trading:} An additional slicing to this problem includes goal specific or goal independent data valuation depending on whether there is a specific well-defined goal for the data purchase or if it is exploratory by design for a goal that is currently undefined; but would be drafted later on. \par \textbf{Horizontal or vertical data acquisition:} In addition to all these situations of data valuation, yet another categorization is based on whether the data acuiqition is being done vertically (in terms of acquiring attributes/columns) or horizontally (acquiring records/rows) as in\citep{ghorbani2019data,jia2019towards}. This terminology of 'vertical partitioning' and 'horizontal partitioning' extends from the databases as well as distributed systems research communities.

\begin{figure}[H]
\centering
  \includegraphics[width=70mm,height=62mm]{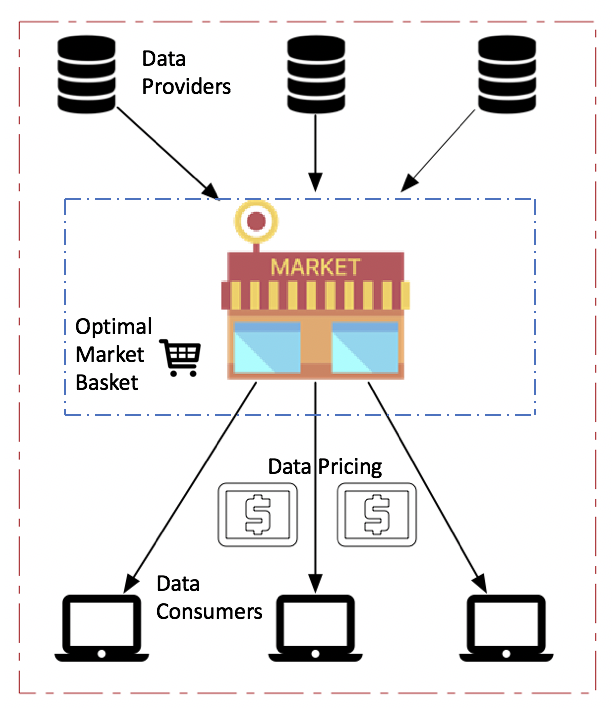}
  \caption{Data market showing data providers, data customers, notions of market basket and data pricing}
  \label{fig:boat0}
\end{figure}

\textbf{Privacy aware data valuation:} Ideally such a data valuation needs to be acheived by looking at as few records per data source as possible or via privacy aware AI \citep{vepakomma2018split,vepakomma2018leakage,vepakomma2018no,gupta2018distributed}. Pooling of all data at a centralized location defeats the central purpose and the data sharing constraints of privacy, security, safety, fairness and resource efficiency need to be kept in consideration with regards to a data valuation solution for data markets. \par
\textbf{Relevance and diversity of data acquisition:} An optimal data purchase under these constraints needs to cater to high utility and low redundancy (high diversity) of data in terms of incremental benefit obtained. There is often a tradeoff of utility vs. diversity of data that needs to be considered in realistic settings. This concept has been the guding principle for techniques like sure independence screening (SIS) and conditional sure independence screening \citep{fan2008sure,zhong2015iterative,barut2016conditional} currently actively being studied in the field of statistics and in min redundancy max relevance (mRMR) \cite{peng2005feature} in the field of data mining, during the precursory periods of current day AI and machine learning. 

As shown in Figure 1, a robust data valuation acts as a good input for data pricing as well as for building an optimal market basket of data for every data consumer. \par
%In Table 1 below we summarize these factors along with a possible sampling of technical solutions for solving the data valuation problem.
\par The intent of sharing these possibilities is to motivate further discussion and research. We summarize some of these points with regards to data valuation in the context of data markets as shown below:
\begin{figure}[H]
\centering
  \includegraphics[width=120mm,height=55mm]{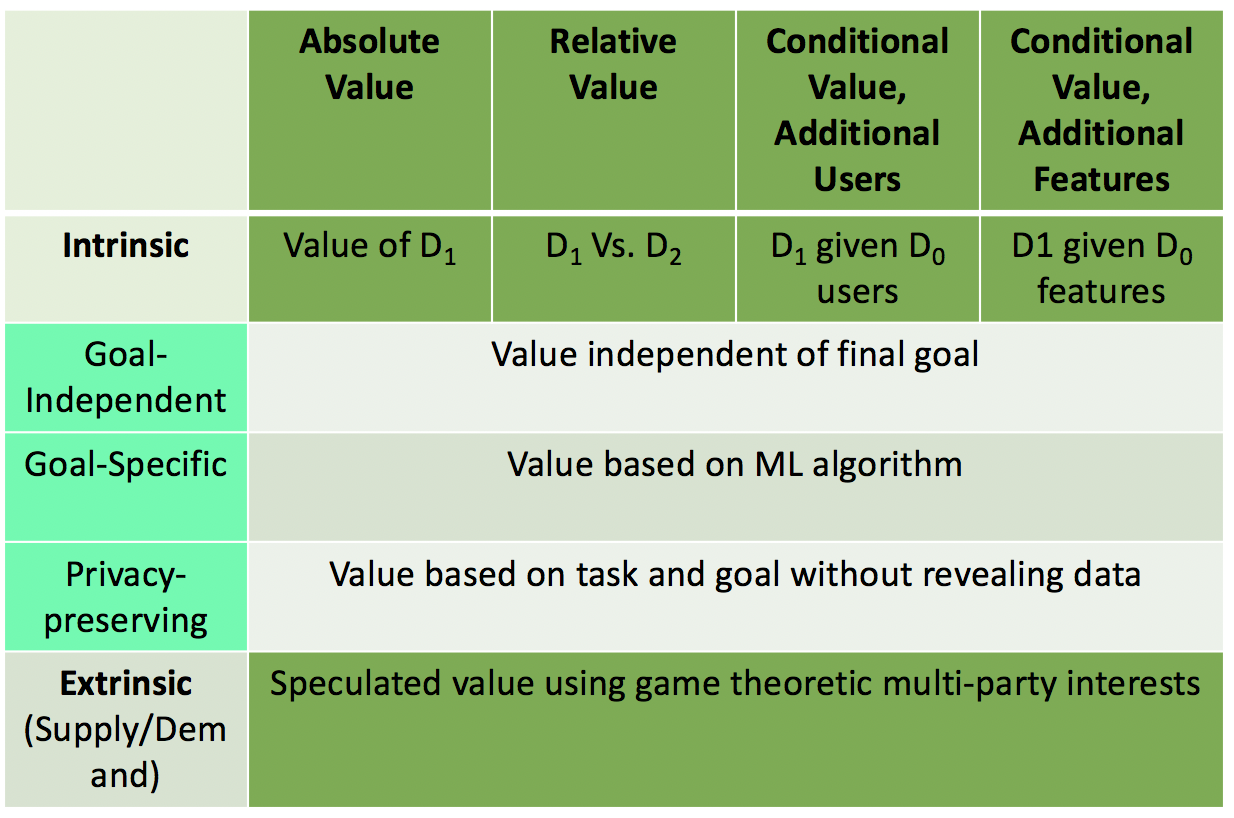}
  \caption{Landscape of data valuation problems for data markets}
  \label{fig:boat0}
\end{figure}

\section{Data is very complex to price}
The value of an incremental unit of data is also conditionally dependent on data already possessed by the prospective data buyer entity that is valuating it. This is because one would like to obtain relevant yet diverse data from what is already available in-house. In addition, data can be acquired for performing either a similar or a more diverse task in comparison to the current use-cases being applied on data that is already available in-house. Also, there are so many archetypes that it is difficult to find a proxy variable (like weight or number in the case of other goods) that can be used to define the data. Since seamless discovery and a small spread in price is essential for a marketplace to function well, it has been challenging thus far to create a functioning data marketplace. A thorough data pricing startegy needs to adhere to the following guiding principles.\\\\
\textbf{Data Pricing Guidelines}
\begin{enumerate}
\item Liquidity: models freshness of dat in terms of value vs diminished/increased value over time
\item Traceability: can be only ‘sold’ once, or sold non-exclusively
\item Consent: maintains privacy of owner, tracks consent over time, and reduces friction with smart contracts or data concierges.
\item Neutrality: accessible to all buyers to prevent unfair trading practices. Otherwise, it would encourage some players (be it large or small) to unfairly price out the rest of prospective buyers during the trading.
\item Recourse: Allows for calling back, provides right to be forgotten, allows for some course correction, broadly remains self-sustaining.

\end{enumerate}

\section{Data sharing challenges that data markets need to address}
Although acquiring the right amount of quality data is ideal, data sharing is heavily impeded by friction caused by lack of trust, data sharing regulations such as HIPAA/GDPR, lack of ease and lack of incentive. We further expand on these factors that cause data friction.

\begin{enumerate}
    \item \textbf{Lack of incentives:} 
     \begin{enumerate}
         \item Large organizations need incentive mechanisms to share data with small players. For example, an incentive for data sharing between large centralized hospitals and local clinics, testing centers could be to foster better provision of health.
\item Big tech players have taken a lead and are rapidly collecting and hoarding data while monopolizing the data resources and are preventing small players from entering into data acquisition. This stifles innovation.
\item Individuals need incentives to share their data as they happen to generate and own tremendous amount of data on a daily basis. But this leads to the burden of consent management which is too complex to manage granularly across different modalities, time horizons, and trust-levels in data buyers.
\item Governments and non-profit are often not allowed to sell data for monetary gains .

     \end{enumerate}
    \item \textbf{Lack of ease of sharing data:}
Due to lack of automated processes, digitization, access to data pre-processing pipelines, compatible data schemas, lack of standardization across data sources and other forms of siloing of socially beneficial data; seamless data sharing is restricted. To summarize, these factors include:
    \begin{enumerate}
        \item Lack of digitization and lack of use cases
\item Lack of data standardization across multiple sources
\item Collection of data currently will likely cost more than market price of data
\item Socially beneficial good data is locked away (e.g. with government, non-profits, hospitals, remote sensing data)

    \end{enumerate}
    \item \textbf{Lack of trust:} Data sharing can also be impeded by the factors of market forces, need for maintaining trade secrets, competitive economy that impedes trust, fear of losing control and accountability over future usage of data for adversarial purposes. To summarize, these factors include cases when:
\begin{enumerate}
    \item Data owner does not trust what the buyer will do with 
data in a competing environment
  \item Data indirectly contains trade secrets of the data owner
  \item Fear of adversarial future usage of shared data
\end{enumerate}
    \item \textbf{Regulations:} Data sharing is regulated for privacy, security, fairness and safety and therefore any data transactions for performing basic data analysis or for any advanced AI/ML usecases has to be aware of these constraints and  be able to safely circumvent these friction points while also maintaining compliance with the law. To summarize: 
    
    \begin{enumerate}
        \item In sectors such as health, finance and cybersecurity that are tightly governed by local, federal and international data sharing regulations such as HIPAA, GDPR, COX, PCI, SHIELD, we need a new strategy for safe data sharing. 
        \item Policies for inter as well as intra organizational data sharing have to be adhered to.
\item The origination of data may have country specific regulations on usage. Therefore international regulations need to be adhered to with respect to both the data provider as well as the data consumer.
\item There are policies where data cannot physically leave the premises of the data owners.

    \end{enumerate}

\section{Governance and encouragement for a data market ecosystem} In addition from the perspective of governance, the following would be key to support the setup as well as to sustain a good ecosystem for data markets \cite{aalekhTalk}.

\begin{enumerate}
    \item Need to support technological solution vs market solution vs policy-driven solutions.
\item Data governance policies by undertaking a study of changes needed in existing legal/regulatory frameworks.
\item Standardization of data sharing
\item Setting up national ‘nodes’ of servers for data exchange (like stock exchanges)
\item ‘Clean Data’ credits like ‘clear air’ carbon credits
\item Treat data as labor (it’s from activity, that creates value)
\item Ethics and bias: self certification as well as audits
\end{enumerate}
\end{enumerate}
%\section{Conclusion}

\printbibliography

\end{document}